\documentclass[11pt, letterpaper]{article}

\usepackage[margin=1in]{geometry}
\usepackage[numbers]{natbib}
\usepackage[reqno]{amsmath}
\usepackage{amssymb,amsthm,graphicx,verbatim,url,verbatim}
\usepackage{color}
\usepackage{url}
\usepackage{rotating}
\usepackage{setspace}
\usepackage{bbm}
\usepackage{wrapfig}
\usepackage{lipsum}
\usepackage{algorithm}
\usepackage{algorithmic}

\usepackage{tabularx} 
\usepackage{booktabs} 

\usepackage[table]{xcolor}
\newcommand{\highlight}[1]{\cellcolor{green!25}#1}

\usepackage[lofdepth,lotdepth]{subfig}

\usepackage{color}\definecolor{spot}{rgb}{0.6,0,0}

\usepackage[pdftex, bookmarksopen=true, bookmarksnumbered=true,
  pdfstartview=FitH, breaklinks=true, urlbordercolor={0 1 0},
  citebordercolor={0 0 1}, colorlinks=true,
            citecolor=spot,
            linkcolor=spot,
            urlcolor=spot,
            pdfauthor={Fitzsimons, Honaker, Shoemate, Singhal},
pdftitle={Title}]{hyperref}

\title{Private Means and the Curious Incident of the Free Lunch}
\author{Jack Fitzsimons\thanks{\texttt{jack@oblivious.com}; Oblivious, \url{https://oblivious.com}.}
\and James Honaker\thanks{\texttt{jhonaker@mozilla.com}; Mozilla Anonym, \url{http://anonymco.com}; \url{http://hona.kr}}
\and Michael Shoemate\thanks{\texttt{shoematem@seas.harvard.edu}. OpenDP, John A. Paulson School Of Engineering And Applied Sciences, Harvard University.}
\and Vikrant Singhal\thanks{\texttt{vikrant@seas.harvard.edu}. OpenDP, John A. Paulson School Of Engineering And Applied Sciences, Harvard University.}}

\begin{document}

    \newtheorem{theorem}{Theorem}
    \newtheorem{proposition}{Proposition}
    \newtheorem{corollary}{Corollary}[theorem]
    \newtheorem{definition}{Definition}
    
    \newtheorem{lemma}[theorem]{Lemma}
    
\newcommand{\EE}{\mathbb{E}}  
\newcommand{\vars}{\mathit{Var}}  
\newcommand{\ex}[2]{{\ifx&#1& \mathbb{E} \else \underset{#1}{\mathbb{E}} \fi \left[#2\right]}}
\newcommand{\pr}[2]{{\ifx&#1& \mathbb{P} \else \underset{#1}{\mathbb{P}} \fi \left[#2\right]}}
\newcommand{\var}[2]{{\ifx&#1& \mathrm{Var} \else \underset{#1}{\mathrm{Var}} \fi \left(#2\right)}}

\newcommand{\SimplexMean}{\mathrm{DPASMean}}
\newcommand{\VarPP}{\mathrm{VarianceRegulator}}

\newcommand{\poly}{\mathrm{poly}}
\newcommand{\polylog}{\mathrm{polylog}}

\newcommand{\id}{\mathbb{I}}  
\newcommand{\Lap}{\mathrm{Lap}}  
\newcommand{\TLap}{\mathrm{TLap}}  
\newcommand{\opt}{\mathrm{opt}}
\newcommand{\OPT}{\mathrm{OPT}}

\newcommand{\X}{\mathcal{X}}  
\newcommand{\Q}{\mathcal{Q}}  
\newcommand{\M}{\mathcal{M}}  

\newcommand{\cA}{\mathcal{A}}
\newcommand{\cB}{\mathcal{B}}
\newcommand{\cC}{\mathcal{C}}
\newcommand{\cD}{\mathcal{D}}
\newcommand{\cE}{\mathcal{E}}
\newcommand{\cF}{\mathcal{F}}
\newcommand{\cG}{\mathcal{G}}
\newcommand{\cH}{\mathcal{H}}
\newcommand{\cI}{\mathcal{I}}
\newcommand{\cJ}{\mathcal{J}}
\newcommand{\cK}{\mathcal{K}}
\newcommand{\cL}{\mathcal{L}}
\newcommand{\cM}{\mathcal{M}}
\newcommand{\cN}{\mathcal{N}}
\newcommand{\cO}{\mathcal{O}}
\newcommand{\cP}{\mathcal{P}}
\newcommand{\cQ}{\mathcal{Q}}
\newcommand{\cR}{\mathcal{R}}
\newcommand{\cS}{\mathcal{S}}
\newcommand{\cT}{\mathcal{T}}
\newcommand{\cU}{\mathcal{U}}
\newcommand{\cV}{\mathcal{V}}
\newcommand{\cW}{\mathcal{W}}
\newcommand{\cX}{\mathcal{X}}
\newcommand{\cY}{\mathcal{Y}}
\newcommand{\cZ}{\mathcal{Z}}

\newcommand{\RR}{\mathbb{R}}  
\newcommand{\NN}{\mathbb{R}}  
\newcommand{\R}{\mathbbm{R}}  
\newcommand{\Z}{\mathbbm{Z}}  
\newcommand{\ONE}{\mathbbm{1}}  

\newcommand{\iprod}[1]{\langle #1 \rangle}  
\newcommand{\llnorm}[1]{\left\lVert#1\right\rVert_2}  
\newcommand{\fnorm}[1]{\left\lVert#1\right\rVert_F}  
\newcommand{\norm}[1]{\left\lVert#1\right\rVert}  
\newcommand{\sgnorm}[1]{\left\lVert#1\right\rVert_{\Psi^2}}  

\newcommand{\Score}{\textsc{Score}}
\newcommand{\Match}{\mathrm{Match}}
\newcommand{\Median}{\mathrm{Median}}
\newcommand{\pcount}{\mathrm{Count}}

\renewcommand{\epsilon}{\varepsilon}  
\renewcommand{\k}{\kappa}  
\newcommand{\eps}{\epsilon}  
\newcommand{\renyi}{R\'enyi\xspace}  
\newcommand{\Renyi}{R\'enyi\xspace}  
\newcommand{\wh}{\widehat}  
\newcommand{\rank}{\mathrm{Rank}}  
\newcommand{\Tr}{\mathbf{Tr}}

\newcommand{\dr}[3]{\mathrm{D}_{#1}\left(#2\middle\|#3\right)}  
\newcommand{\stcomp}[1]{\overline{#1}}  
\newcommand{\abs}[1]{\left| #1 \right|}  
\newcommand{\chisq}[1]{\chi^2\left( #1 \right)}  
\newcommand{\ball}[2]{\mathit{B}_{#2}\left( #1 \right)}  

\newcommand{\sidecomment}[1]{\hfill\texttt{\textbf{// #1}} \\}  
\newcommand{\fix}{\marginpar{FIX}}
\newcommand{\new}{\marginpar{NEW}}

\newcommand{\eg}{\textit{e.g\@.}}  
\newcommand{\etc}{\textit{etc.}}  
\newcommand{\etal}{\textit{et~al\@.}}  
\newcommand{\ie}{\textit{i.e\@.}}  

\newcommand{\Argmax}{Argmax}  
\newcommand{\GAPMAX}{\mathrm{GAP\text{-}MAX}}  
\newcommand{\Sym}{\mathsf{Sym}}  
\newcommand{\myeqand}{\quad \textrm{and} \quad}  

\newcommand{\GGNmax}{GGNmax}  
\newcommand{\Laplace}{Laplace}  
\newcommand{\Gaussian}{Gaussian}  

\newtheorem{remark}[theorem]{Remark}

\allowdisplaybreaks

\maketitle

\vspace{-0.7cm}
\abstract{We show that the most well-known and fundamental building blocks of DP implementations -- sum, mean, count (and many other linear queries) -- can be released with substantially reduced noise for the same privacy guarantee.  We achieve this by projecting individual data with worst-case sensitivity $R$ onto a simplex where all data now has a constant norm $R$.  In this simplex, additional ``free'' queries can be run that are already covered by the privacy-loss of the original budgeted query, and which algebraically give additional estimates of counts or sums.}

\section{Introduction}

As frameworks for differential privacy (DP)~\cite{DworkMNS06} grow in popularity, some of the most common base operations in production systems remain the calculation of primitive linear statistics such as sums, counts, and means of data.  These operations have largely been considered closed problems since their introduction -- they are the opening examples in any pedagogical material, the fundamental building blocks of most DP algorithms, and the backbone of most practical implementations. Despite that, we show there is still room for improvement in the estimation error in practical deployments.

In this work, we focus primarily on the privacy-preserving release of means, but whilst relating the problems of computing the sums of our data and the counts. We consider the setting where the total number of elements in the dataset is unknown. When the dataset's size is known, the most popular approach to finding the mean is simply to calculate the differentially private sum of values and divide it by the known set size, as introduced in~\cite{DworkMNS06}. This simplified scenario is leveraged in many systems today, including the popular Differential Privacy Library (DiffPrivLib)~\cite{HolohanBAL19} and Opacus~\cite{YousefpourSSTPMNGBZCM21}.
The scenario that we tackle is most commonly calculated with two privacy-preserving queries: the sum and the count. The mean estimate is then post-processed from the privacy-preserving (``noisy'') sum divided by the noisy count as seen in, for example, OpenDP~\cite{OpenDPLibrary}, SmartNoise SQL~\cite{Smartnoise}, Qrlew~\cite{Qrlew}, and PipelineDP~\cite{PipelineDP}. These, in turn, form the basis of composite systems such as PySyft~\cite{PySyft} and Antigranular~\cite{Antigranular}.

We will show that in our unknown-dataset-size setting, enough information is being left on the table, which would consume no additional privacy budget to leverage and would lead to further improvements in the estimation error. Specifically, we will show that by constructing vector responses with known sensitivity, we can always lower our estimate's variance, typically by a half. 

\subsection{Related Work}

There has been a lot of work in recent years on differentially private mean estimation.
Mean estimation is one of the most fundamental question in statistics, enjoying significant attention from the perspective of DP (e.g.,~\cite{KuleszaSW24, Lebeda24, KarwaV18, BunS19, KamathLSU19, KamathSU20, WangXDX20, DuFMBG20, BiswasDKU20, CaiWZ21, BrownGSUZ21, HuangLY21, LiuKKO21, LiuKO22, KamathLZ22, HopkinsKM22, KothariMV22, TsfadiaCKMS22, DuchiHK23, CovingtonHHK21, NikolovT23, KamathMRSSU23, AumullerLNP24, Singhal24}).
Another direction of interest is ensuring privacy when one individual is allowed to contribute multiple data points \cite{AgarwalKMMSU25, LiuSYKR20, LevySAKKMS21, GeorgeRST22, RameshwarT25} (also known as, \emph{user-level differential privacy}). Finally, mean estimation with access to small amounts of public data~\cite{BieKS22} is also a new upcoming direction.
We refer the reader to \cite{KamathU20} for more details on the recent work on private statistical estimation. Finally, a private selection of the clipping bounds parameter is often critical in applied settings, which recent work explores, for example in~\cite{KarwaV18, KamathLSU19, BiswasDKU20, CovingtonHHK21, Durfee24, KamathMSSU22, AshtianiL22, KothariMV22, TsfadiaCKMS22}, and we recognize this as a central issue, but take the bounds to be known \emph{a priori} or already privately released.

One important set of results from the aforementioned list that we would like to highlight is from the independent work of~\cite{KuleszaSW24}. They study the problem of mean estimation of univariate data under the add/remove model of DP when the size of the data set is unknown (or what we call, ``unbounded DP'') and happen to have a very similar algorithm as ours. They also prove lower bounds showing the optimality of the algorithm. We additionally show that this approach also works for computing weighted means and is compatible with the Gaussian mechanism as it preserves $\ell_2$ sensitivity. 

The work of~\cite{Lebeda24} also bears a resemblance due to its geometric intuition, but is distinct: the work exploits the sensitivity space induced by vector-valued queries whose elements differ in the same direction, in a manner similar to monotonic scoring functions for the exponential mechanism. Augmenting queries, as discussed in our work, preserves the structure of the sensitivity space, thus allowing the use of the correlated Gaussian mechanism from~\cite{Lebeda24}.

Relevant to our work, \cite[Appendix~D]{KamathMRSSU23} provide algorithms to output unbiased means of distributions coming from certain families when the dataset size is unknown under both pure and approximate DP. Although we do not focus on unbiasedness, we focus on obtaining the optimal mean-squared error in the non-distributional setting. As a general point of art, that within a workload of statistics, there may be many DP released answers to the same query and they should be combined for statistical efficiency we see this as connected to the post-processing of DP trees~\cite{hay2009boosting, honaker2015efficient}.

\subsection{Contributions}

In this work, we assume that datasets differ in the addition or removal of rows (unbounded DP).  We wish to learn the mean of a variable that is bounded $[0,R]$ (or can be transformed to such), where $R$ is known or already released by another DP mechanism, and consider loss in the estimate to be captured by mean squared error (MSE). We provide an algorithm that augments each data point, $x$, with its additive complement, $R-x$. Then we output the $2$-dimensional vector of the sums of the two columns privately. Without privacy, if we add the two sums together, we will get the count (or the size) of the dataset. Adding the two privatized versions, we get a ``free'' count of the dataset privately, as well. The key observation here is that augmenting the data with another column this way does not increase the sensitivity of the $2$-dimensional sum, which helps us get additional information privately ``for free''. We provide experimental results to compare the utility of our algorithm with that of other known DP methods. We also extend this technique to compute weighted means of the data.

\section{Preliminaries}

We first define two well-known forms of differential privacy. Two datasets are considered to be ``neighboring'' if they differ on at most one row.

\begin{definition}[Differential Privacy (DP)~\cite{DworkMNS06}]
    A randomized algorithm $M: \cX^n \rightarrow \cY$ satisfies \emph{$(\eps,\delta)$-differential privacy ($(\eps,\delta)$-DP)} if for every pair of neighboring datasets $X, X' \in \cX^n$,
    $$
        \forall Y \subseteq \cY~~~\pr{}{M(X) \in Y} \leq e^{\eps} \pr{}{M(X') \in Y} + \delta.
    $$
\end{definition}

\begin{definition}[Concentrated Differential Privacy (zCDP)~\cite{BunS16}]
    A randomized algorithm $M: \cX^n \rightarrow \cY$
    satisfies \emph{$\rho$-zCDP} if for
    every pair of neighboring datasets $X, X' \in \cX^n$,
    $$\forall \alpha \in (1,\infty)~~~D_\alpha\left(M(X)||M(X')\right) \leq \rho\alpha,$$
    where $D_\alpha\left(M(X)||M(X')\right)$ is the
    $\alpha$-R\'enyi divergence between $M(X)$ and
    $M(X')$.\footnote{Given two probability distributions
    $P,Q$ over $\Omega$,
    $D_{\alpha}(P\|Q) = \frac{1}{\alpha - 1}
    \log\left( \sum_{x} P(x)^{\alpha} Q(x)^{1-\alpha}\right)$.}
\end{definition}

These definitions are closed under post-processing of the private outputs.

\begin{lemma}[Post-Processing~\cite{DworkMNS06, BunS16}]\label{lem:postprocessing}
    If $M : \cX^n \to \cY$ is $(\eps,\delta)$-DP (or $\rho$-zCDP) and
    $P : \cY \to \cZ$ is any randomized function, then
    the algorithm $P \circ M$ is $(\eps,\delta)$-DP (or $\rho$-zCDP).
\end{lemma}

Next, we define a useful DP primitive algorithm -- the Gaussian mechanism.

\begin{definition}[$\ell_2$-Sensitivity]
Let $f : \cX^n \to \RR^d$ be a function, its \emph{$\ell_2$-sensitivity} is
$$
\Delta_{f} = \max_{X \sim X' \in \cX^n} \| f(X) - f(X') \|_{2}
$$
\end{definition}

\begin{lemma}[Gaussian Mechanism]\label{lem:gaussiandp}
    Let $f : \cX^n \to \RR^d$ be a function
    with $\ell_2$-sensitivity $\Delta_{f}$.
    Then the Gaussian mechanism:
    \begin{enumerate}
        \item $M(X) = f(X) + \cN\left(0,\tfrac{2 \Delta_{f}^2
        \ln(2/\delta)}{\eps^2} \cdot \id_{d \times d}\right)$
    satisfies $(\eps,\delta)$-DP;
    \item $M(X) = f(X) + \cN\left(0,\tfrac{\Delta_{f}^2}{2\rho} \cdot \id_{d \times d}\right)$
    satisfies $\rho$-zCDP.
    \end{enumerate}
\end{lemma}

\section{Warm Up: Traditional Mean Estimation}\label{s:trad} 

Typically, the mean for a known dataset size, $N$, is computed by taking a DP summation of the data points and dividing by the dataset size,
\begin{align}
\hat{x} &= \frac{(\sum_{i=1}^N x_i) + Z}{N},
\end{align}
where $Z$ is noise added by either the Gaussian or Laplacian distribution with appropriate parameterization depending on the form of privacy guarantee desired. In the scenario where the dataset size is unknown, we additionally release the dataset size in a privacy-preserving way and substitute this in for the private value, as:
\[
\hat{N} = \left(\sum_{i=1}^N 1\right) + W, \hspace{10mm}
\hat{x} = \frac{(\sum_{i=1}^N x_i) + Z}{\hat{N}}
\]
Here, there are two independent sources of error: the noisy count ($Z$) and the noisy sum ($W$).  

\section{Simplex Augmentation Transformation} 

Assume we need the mean of user values $X: x_i \in [0,R]$, where the total number of users, $N$, is also private information. We are going to augment the values in $x_i$ by projecting them onto the simplex (the space of non-negative vectors that sum to 1).\\ 

\noindent \begin{minipage}{0.55\textwidth}
We first expand $X$ into the 2-vector $Y$ as:
\begin{align}
y_i = (x_i, R - x_i)
\end{align}
and compute the (private) sum on both columns:
\begin{align}
s_1 &= \sum_{i=1}^N y_{i1} = \sum_{i=1}^N x_i; \\
s_2 &= \sum_{i=1}^N y_{i2} = \sum_{i=1}^N R-x_i = NR -\sum_{i=1}^N x_i = NR - s_1.
\end{align}
For all $i$, $y_i$ has $\ell_1$ norm $R$, thus for $\varepsilon$-DP using the Laplacian mechanism, the average privacy loss and the upper bound on the privacy loss are equal. As such, we can say that a query on $(s_1,s_2)$ is optimal in terms of the privacy-utility trade-off.
\end{minipage}
\hspace{0.3cm}
\begin{minipage}{0.3\textwidth}
\vspace{-0.5cm}
\includegraphics[width=\textwidth]{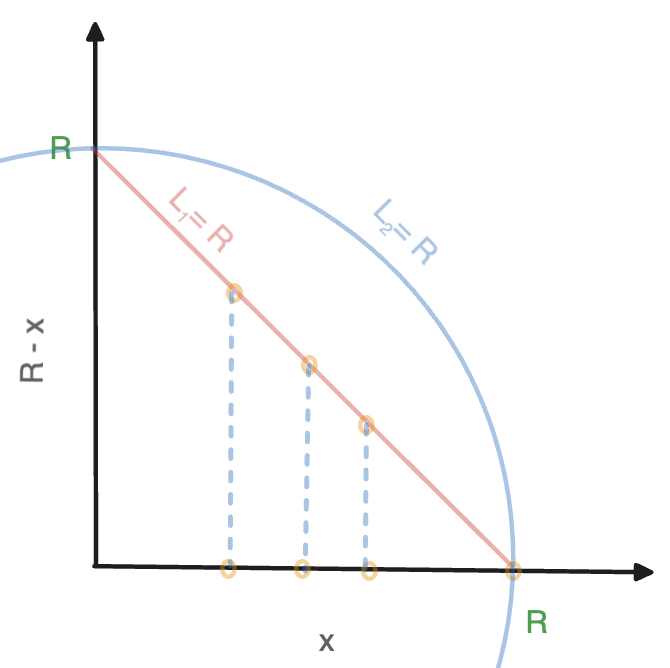}\\
\textbf{Figure 1:}\emph{ The projection of $x$ onto the 2-D simplex to form $y$.} 
\setcounter{figure}{1}  
\end{minipage}\\

 Alternatively, if we are interested in zero-concentrated DP (zCDP)~\cite{BunS16}, by the triangle inequality the $\ell_2$ norm of $y_i$ is at most $R$, and we can still release both sums of both columns with the same budget.

\subsection{Free Counts under Unknown Dataset Size}

Under either privacy-loss definition, for total privacy budget $\rho$ (or $\varepsilon$), we can release both sums with the Gaussian (or the Laplace) mechanism simultaneously as:
\begin{align}
m_1 &= s_1 + Z_1; \qquad Z_1 \sim \mathcal{N}\left(0,\frac{R^2}{2\rho}\right), \qquad m_2 = s_2 + Z_2; \qquad Z_2\sim\mathcal{N}\left(0,\frac{R^2}{2\rho}\right). \label{eq:m2}
\end{align}

The DP release $m_1$ is the original sum that we were previously releasing at the same $\rho$ privacy budget;  there is no decrease in the utility on the sum.  However, we can also compute (by post-processing) a DP estimate of the user count as:
\begin{align}
    \hat{N} = \frac{m_1 + m_2}{R}= \frac{s_1 + Z_1 + NR - s_1 + Z_2}{R} = N + \frac{Z_1 + Z_2}{R},
\end{align}
where the right term is a mean-zero Gaussian random variable with variance:
\begin{align}
    \textrm{Var}\left(\frac{Z_1 + Z_2}{R}\right)= \frac{1}{R^2}\textrm{Var}(Z_1 + Z_2) = \frac{1}{R^2}\cdot\frac{2R^2}{2\rho} = \frac{1}{\rho}. \label{eq:count}
\end{align}
Note that this DP estimate of $N$ is obtained for ``free'' in the sense that no additional budget is spent beyond what had been originally allocated for the sum, and with no degradation in the utility of the sum for this additional release.  At first, this seems like an impossible trick.  The intuition is that to release the sum we are already assuming that each user may have data with sensitivity $R$, but in actuality, most users do not -- by moving user data on to the 2D-simplex we can utilize this sensitivity that we have already budgeted against.We state this algorithm formally as Algorithm~\ref{alg:simplex-mean-unknown}.

\begin{algorithm}
\caption{DP Mean Estimator with Simplex Augmentation Unknown Size $\mathrm{DPASMean}_{\rho,R}(X)$}\label{alg:simplex-mean-unknown}
\begin{algorithmic}[1]
    \REQUIRE Samples $X_1,\dots,X_N \in \mathbb{R}$. Parameters $\rho>0$ and $R>0$, such that $|X_i| \leq R$ for all $i$.
    \ENSURE Mean of $X$.
    \STATE Expand $X$ into a $2$-vector as $y_i=(x_i,R-x_i)$.
    \STATE Set $s_1 \gets \sum\limits_{i=1}^{N}{y_{i1}}$ and $s_2 \gets \sum\limits_{i=1}^{N}{y_{i2}}$.
    \STATE Set $m_1 \gets s_1 + Z_1$ and $m_2 \gets s_2 + Z_2$, where $Z_1,Z_2 \sim \mathcal{N}\left(0,\tfrac{R^2}{2\rho}\right)$.
    \STATE $\hat{N} \gets \tfrac{m_1+m_2}{R}$.
    \STATE \textbf{Return} $\hat{\mu} \gets \tfrac{s_1}{\hat{N}}$.
\end{algorithmic}
\end{algorithm}

\begin{theorem}\label{thm:simplex-dp}
    Algorithm~\ref{alg:simplex-mean-unknown} satisfies $\rho$-zCDP.
\end{theorem}
\begin{proof}
    The $\ell_2$ norm of each row of $Y$ is upper-bounded by $R$. Therefore, adding Gaussian noise scaled to $R$ and $\rho$ is sufficient to guarantee zCDP (Lemma~\ref{lem:gaussiandp}).
\end{proof}

\subsubsection{Example} As a point of comparison, assume (as is typical) we were to divide the privacy budget into two halves to individually estimate the sum of the first column and the size of the dataset.  In this cas, the variance of the sum increases by a factor of $2$ (over what is seen in (\ref{eq:m2})) because of the reduction in $\rho$ for the sum, while the count/size of the dataset has exactly the same variance as in Equation~\ref{eq:count}, even though it received half the budget, rather than being obtained as a free by-product.

\subsubsection{Improving the Error in the Count}\label{ss:improv_counts}

Although the count comes with a fixed error proportional to the budget on the mean, if we are in a setting where the resulting utility of $\hat{N}$ is not sufficient, we can allocate additional privacy budget to an additional DP release of the count.  We then weight together our answers to improve our estimate.  Explicitly, if we spend additional privacy-loss budget $\rho’$, then we can release:
\begin{align}
    N’ = N + W; \qquad W \sim \mathcal{N}\left(0,\frac{1}{2\rho’}\right),
\end{align}
from which we can derive an optimal estimate, $N^*$, by weighting together our two DP estimates as:
\begin{align}
    N^* = w\hat{N} + (1-w)N’,
\end{align}
where the weight $w$ comes from inverse variance weighting as:
\begin{align}
    w = \frac{\textrm{Var}(\hat{N})^{-1}}{\textrm{Var}(\hat{N})^{-1} + \textrm{Var}(N’)^{-1}} = \frac{\rho}{\rho + 2\rho’}.
\end{align}
Inverse variance weighting is statistically efficient -- meaning it gives the lowest variance unbiased estimator -- for linear combinations of independent measures. For a demonstration in the context of combining DP releases, see \cite[Remark~1]{honaker2015efficient}.

\subsection{Extension to Weighted Means}
In many real-world applications, not all data points contribute equally to the final statistic. For example, in survey analysis, responses may be weighted by demographic factors, or in financial calculations, transactions may be weighted by their monetary value. Our simplex transformation naturally extends to handle these weighted scenarios.

For a dataset where each point $x_i$ has an associated weight $w_i>0$, we modify our transformation as:
\begin{align}
y_i = (w_i x_i, R - w_i x_i)
\end{align}
where $R$ is either the clamping bound of $\langle w_i x_i \rangle$ or the product of the clamping bounds of $\langle w_i \rangle$ and $\langle x_i \rangle$, depending on how the data has been clamped. The choice between these approaches depends on whether weights and values are clamped independently or jointly.

This weighted transformation preserves all the key properties of our original method:
\begin{itemize}
    \item The $\ell_1$ norm of each transformed point remains constant at $R$
    \item The privacy guarantees hold with the same budget
    \item The utility gains from complementary statistics remain available
\end{itemize}

A practical example arises in demographic surveys where certain population segments are oversampled. If a minority group comprising 10\% of the population represents 30\% of survey responses, we might use weights of $w_i=1/3$ for oversampled responses to restore population representativeness while maintaining privacy guarantees.

\section{Empirical Findings}\label{a:empirics}

We consider three other common techniques (that we describe below: plugin, centered mean, and resize transformation) for means in the setting where dataset size is unknown. For a comparison of the relative performance of these and our new method, we create a distribution of draws for a fixed dataset with a fixed privacy-loss parameter. In these implementations, when necessary, the privacy budget was distributed to the numerator and denominator to equalize their variances, which is empirically where the variance is minimized under those approaches.

\begin{figure}
\begin{center}
\includegraphics[width=0.7\textwidth]{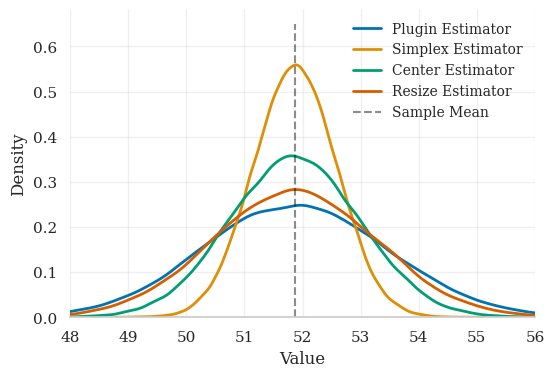}
\end{center}
\caption{Simulations of noise distribution for different mean release algorithms with same privacy-loss guarantee.  Our simplex method has uniformly lowest variance across data distributions. Here we show reduced variance in the release for means of 100 uniformly random data points in $[0, 100]$ using $\rho = 0.5$. See Appendix~\ref{a:empirics} for more details and further analysis.}\label{fig:initial-comp}
\end{figure}

\hspace{0.3cm} Across these simulations, the simplex estimator provides the lowest variance (Figure~\ref{fig:initial-comp}). We give a deeper dive into our empirical results in Figure~\ref{fig:performance}, but the improvements of our method are quite stark and constant. None of these other approaches are data dependent, so we do not see these findings changing in other contexts.

\subsection{Plugin}

The first is where the sum and count are individually budgeted and released and then substituted into the mean formula (``plugged-in'' in the statistical estimator sense), as described in Section~\ref{s:trad}.

\subsection{Centered Mean}\label{a:other}


Assume you want to compute the sum of each column in a rectangular array $x$ with $d$ columns and $n$ observations, subject to the constraint that the $p$-distance of each row from origin $O$ (a vector) is at most $R$.
The sensitivity of the sum is $\Vert O \Vert_p + R$. 
Equivalently, in the 1-d case: 
\begin{equation}
\Vert O \Vert_p + R = |O| + R = \max(O, -O) + R = \max(R - O, R + O) = \max(|L|, U),
\end{equation}
where $L = O - R$ and $U = O + R$.

The sensitivity is minimized when $O$ is at zero.
Center the data around zero by subtracting $O$ from each row in $x$ (a 1-stable row-by-row transformation). Compute a DP sum of the centered data with sensitivity $R$, denoted $m_i$, and a DP count, denoted $n$.
An unbiased estimate of the DP sum of the original data is $n \cdot O + m$.
The variance of this estimate can be up to halved compared to a direct estimate, and the sum (centered around zero) becomes more numerically stable~\cite{LiLSY17}.

\subsection{Resize Transformation}
One issue with private means with private counts is the numerator is a sum of N objects while the denominator has a privatized noisy count that disagrees.  Resizing is a technique where first a noisy count of the dataset is constructed.  If that count is too low, the original dataset can be randomly downsampled without loss of privacy to that count.  If the count is too high, new observations can be imputed (either a prior belief of the mean, or uniformly at random across the bounds of the data).  Then a privatized sum is computed.  The advantage is that the number of elements in that sum is always agrees with the denominator used to create the mean.  Thanks to Christian Covington for identifying this approach~\cite{Covington20}.

\subsection{Detailed Results}

In order to demonstrate the effectiveness of the simplex algorithm in comparison to standard estimates, we generate 100 data points at random from:

\begin{itemize}
    \item A Log Normal distribution with location 0, scale 1.
    \item A Normal distribution with zero mean and unit variance.
    \item A Uniform distribution ranging 0 to 100. 
\end{itemize}

We compare the mean estimators on each of these stylized datasets using both the Gaussian mechanism (with $\rho = 0.5$) and the Laplacian mechanism (with $\varepsilon = 0.5$). We recorded the distribution of estimates of 10,000 releases per estimator. We recorded the performance of each of the estimators in terms of the absolute error and root mean squared error (RMSE). In all examples, the proposed simplex algorithm far exceeds the performance of the others as seen in Table~\ref{tab:performance}.

\begin{table*}[h!]
\centering
\begin{tabular}{@{}lcccccc@{}}
\toprule
& \multicolumn{2}{c}{Log Normal} & \multicolumn{2}{c}{Normal} & \multicolumn{2}{c}{Uniform} \\
\cmidrule(lr){2-3} \cmidrule(lr){4-5} \cmidrule(lr){6-7}
Estimator & Laplace & Gaussian & Laplace & Gaussian & Laplace & Gaussian \\
\midrule
\textbf{Avg Abs Err} & &  &  &  &  &  \\
Plugin  & 0.3995 & 0.1159 & 0.2843 & 0.0860 & 4.6319 & 1.2742 \\
Simplex  & \highlight{0.1912} & \highlight{0.0764} & \highlight{0.1501} & \highlight{0.0564} & \highlight{1.5106} & \highlight{0.5689} \\
Center  & 0.4038 & 0.1144 & 0.2828 & 0.0859 & 2.9944 & 0.8908 \\
Resize  & 0.3947 & 0.1126 & 0.2850 & 0.0870 & 4.0127 & 1.1322 \\
\midrule
\textbf{RMSE} & &  &  &  &  &  \\
Plugin & 0.5546 & 0.1451 & 0.4058 & 0.1079 & 6.3217 & 1.5960 \\
Simplex & \highlight{0.2607} & \highlight{0.0953} & \highlight{0.2007} & \highlight{0.0707} & \highlight{2.0225} & \highlight{0.7125} \\
Center & 0.5588 & 0.1440 & 0.4007 & 0.1077 & 4.2435 & 1.1162 \\
Resize & 0.5507 & 0.1413 & 0.4051 & 0.1090 & 5.6646 & 1.4189 \\
\bottomrule
\end{tabular}
\caption{Average Absolute Errors and RMSE for different estimators on Log Normal, Normal, and Uniform distributions with Laplace and Gaussian mechanisms.}\label{tab:performance}
\end{table*}

\begin{figure*}[ht!]
    \begin{center}
    \includegraphics[width=6in]{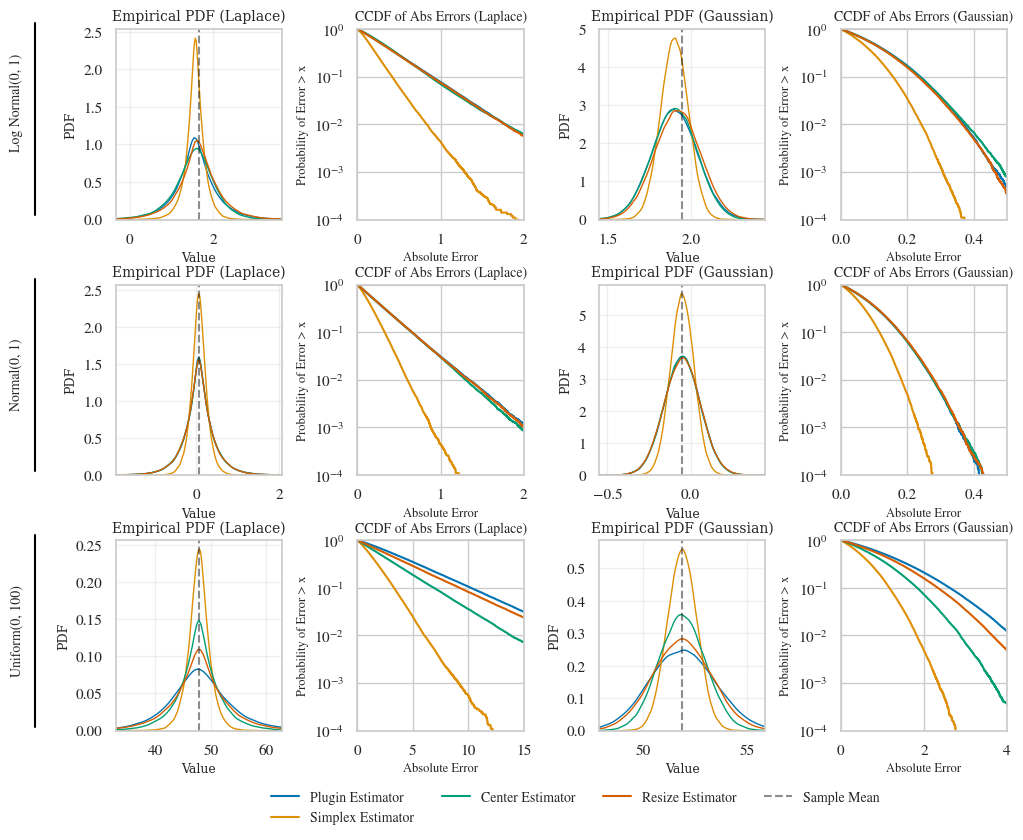}
    \end{center}
    \caption{Estimator performance across three different distributions (displayed in descending order): Log Normal (location 0, scale 0), Normal (mean 0, variance 1), and Uniform (in range 0, 100), each with 100 randomly generated data points. Empirical probability density functions and complimentary cumulative density functions for the Gaussian mechanism ($\rho=0.5$) and Laplacian mechanism ($\varepsilon=0.5$) displayed. The simplex estimator performs the most accurate in every case.}
    \label{fig:performance}
\end{figure*}

Figure~\ref{fig:performance} further demonstrates the performance enhancement. For each estimator and each dataset pairing, the estimates were accumulated to visualise the empirical probability density function versus the true sample mean and the empirical complimentary cumulative density function of the absolute errors.

\section{Conclusion}
Given the set of queries already posed to a dataset, there may be an orthogonal set of questions in which each possible data point has an inverse contribution to the output. As such, one can get the answer to these queries with no additional privacy loss.

\subsection*{Acknowledgements}

The authors would like to thank Christian Covington, David Durfee, Christian Lebeda, Kevin Liou, Rong Xia, and Robert Pisarczyk for their feedback and input. 

\bibliographystyle{alpha}
\bibliography{biblio}

\end{document}